\documentclass[a4paper,11pt]{article}

\pdfoutput=1 % if your are submitting a pdflatex (i.e. if you have
\usepackage{jheppub} % for details on the use of the package, please
\usepackage{amsmath}
\usepackage{graphicx}
\usepackage{dcolumn}
\usepackage{bm}
\usepackage{bbm}
\usepackage{epsfig}
\usepackage{slashed}
\usepackage{amssymb}
\usepackage{color}

\makeatletter

\newcommand{\newc}{\newcommand}
\newc{\renewc}{\renewcommand}

\def\beq{\begin{equation}}
\def\eeq{\end{equation}}
\def\bea{\begin{eqnarray}}
\def\eea{\end{eqnarray}}
\def\bitem{\begin{itemize}}
\def\eitem{\end{itemize}}
\def\ba{\begin{array}}
\def\ea{\end{array}}
\def\bal{\begin{align}}
\def\eal{\end{align}}
\def\bi{\begin{itemize}}
\def\ei{\end{itemize}}
\def\lsim{\mathrel{\rlap{\lower4pt\hbox{\hskip1pt$\sim$}}
    \raise1pt\hbox{$<$}}}         %less than or approx. symbolcoupl
\def\gsim{\mathrel{\rlap{\lower4pt\hbox{\hskip1pt$\sim$}}
    \raise1pt\hbox{$>$}}}

\def\fbi{{\rm {fb^{-1}}}}

\title{$126$ GeV Higgs in Next-to-Minimal Universal Extra Dimensions}

\author[a]{Thomas Flacke,}
\author[b]{Kyoungchul Kong,}
\author[c]{Seong Chan Park}

\affiliation[a]{Department of Physics, Korea Advanced Institute of Science and Technology, \\
335 Gwahak-ro, Yuseong-gu, Daejeon 305-701, Korea}
\affiliation[b]{Department of Physics and Astronomy, University of Kansas, Lawrence, KS 66045, USA}
\affiliation[c]{Department of Physics, Sungkyunkwan University, Suwon 440-746, Korea}

\emailAdd{flacke@kaist.ac.kr}
\emailAdd{kckong@ku.edu}
\emailAdd{s.park@skku.edu}

\abstract{
Discovery of a Higgs boson and precise measurements of its properties open a new window to test physics beyond the standard model. 
Models with Universal Extra Dimensions are not exception. 
Kaluza-Klein excitations of the standard model particles contribute to the production and decay of the Higgs boson.
In particular, the parameters associated with third generation quarks are constrained by Higgs data, which are relatively insensitive to other searches often involving light quarks and leptons. We investigate implications of the 126 GeV Higgs in Next-to-Minimal Universal Extra Dimensions, and  show that boundary terms and bulk masses allow a lower compactification scale as compared to in Minimal Universal Extra Dimensions.
}


\begin{document} 
\maketitle
\flushbottom

%%%%%%%%%%%%%%%%%%%%%%%%%%%%%%%%%%%%%%%%%%%%%%%%%%%%%%%
%%%%%%%%%%%%%%%%       Intro                 %%%%%%%%%%%%%%%%%%%%%%%%%%%%%%
%%%%%%%%%%%%%%%%%%%%%%%%%%%%%%%%%%%%%%%%%%%%%%%%%%%%%%%

\section{Introduction}
\label{sec:intro}

Recent discovery of a Higgs-like boson at the LHC and measurements of its properties open a new window for physics beyond the Standard Model (SM).  Both ATLAS  and CMS collaborations have measured its mass with precision better than that in the top quark mass ($0.5\%$): 
\begin{eqnarray}
m_H = \left \{ \begin{array}{ll}
125.5\pm 0.2 ^{+0.5}_{-0.6} ~{\rm GeV}, &  {\text{ATLAS ($0.43\%$ precision )\cite{ATLASsumref}}},\\
125.7\pm 0.3 \pm 0.3 ~{\rm GeV}, & {\text{CMS ($0.34\%$ precision)\cite{CMSsumref}}}.
\end{array} \right. \label{mass}
\end{eqnarray}
The measured properties of the boson is consistent with the standard model expectation, which is often parameterized by $\mu =\sigma/\sigma_{\rm SM}$ the ratio between the SM expectation and the measured value:  
\begin{eqnarray}
\mu =
\left \{ \begin{array}{ll}
1.30\pm 0.20 & \text{ATLAS \cite{ATLASsumref}},\\
0.80\pm 0.14 & \text{CMS \cite{CMSsumref}}.
\end{array} \right. \label{mu}
\end{eqnarray}

We regard that the discovered boson is actually the Higgs boson in the SM and try to set bounds on new physics model 
comparing the measured data and the expected deviation from new physics.  In general, the radiative production of the Higgs boson through the gluon fusion and its decay to a pair of photon are subject to modification by heavy new colored particles, namely the `top partner' ($t'$) \cite{Kong:2007uu} and electrically charged particles, namely charged gauge bosons and heavy leptons ($W'$ and $\ell' $).  Any new physics model which contains such new particles affects the Higgs physics and can be probed by close examination of  the Higgs data.

In models with universal extra dimensions (UED) \cite{Appelquist:2000nn}, all the standard model particles have their Kaluza-Klein (KK) excitations, 
including new colored particles and electrically charged particles. 
Among them, the KK excitations of the top quark yield significant corrections to the gluon fusion process
due to the largest Yukawa coupling. 
As the fermionic degrees of freedom in models with extra dimensions are doubled, the KK top quark contribution is also enhanced by a factor of 2. 
Also, both the KK $W$ bosons and the KK top quarks contribute to the 1-loop induced decay rate to the diphoton. Even though 1-loop suppressed, the diphoton channel has been regarded as one of a golden channel. 
Other decay channels, which are allowed at tree level, are less significantly modified by the KK states so that we may neglect the effects here. 
There are existing studies on the Higgs production and decay rates in the minimal UED (MUED) model \cite{Petriello:2002uu, Belanger:2012mc} as well as various 5D \cite{dey} and 6D extensions \cite{uedhiggs}.

In this paper, we extend the previous studies by including the effects of bulk mass parameters \cite{Park:2009cs,Huang:2012kz,Kong:2010xk}
and boundary localized terms \cite{CTW,BLKTrefs} following the philosophy of a recent paper \cite{Flacke:2013pla}. 
In Ref. \cite{Flacke:2013pla} it was shown that bulk masses are strongly constrained for leptons and the first two quark families. 
Furthermore, non-uniform boundary terms and bulk masses for leptons, and first and second family quarks typically imply 
large flavor changing neutral current and are thus strongly constrained \cite{Csaki:2010az}. As an important exception, a common boundary parameter for all UED fields is not constrained as it does not induce KK-number violating interactions and only shifts the overall mass scale of the $n$-th Kaluza-Klein mode excitations away from $n/R$, with $R$ being the compactification radius of the extra dimension.

Constraints on parameters associated with third generation quarks are much weaker, and 
their phenomenological implications are very different from those with first and second generations. 
They are particularly important in physics dominated by one-loop corrections, where the large Yukawa coupling of the top plays a crucial role. This applies to electroweak precision tests as well as to Higgs production and decay. We therefore focus on the third generation in the quark sector and consider a UED setup with a common boundary kinetic parameter $r_g$ for the gauge and lepton sector and the first two families (and no bulk masses for those).
We allow for a non-zero bulk mass ($\mu_t \neq 0$) and a different boundary parameter $r_t$ for the third generation. 
This choice leaves us with $r_g,r_t,\mu_t$ and $R^{-1}$ as parameters to be constrained. 

The paper is organized as follows. In the next section, we briefly introduce the Next-to-Minimal UED (NMUED) with bulk mass parameters and boundary localized terms. 
The KK spectra and couplings are collected for the one-loop calculation of the gluon fusion process and radiative decay processes in Section \ref{sec:higgs} and the electroweak bounds are considered in Section \ref{sec:stu}. We show our results in Section \ref{sec:result} taking the latest experimental results into account.

%%%%%%%%%%%%%%%%%%%%%%%%%%
\section{Next-to-Minimal Universal Extra Dimensions}
\label{sec:model}

%%%%%%%%%%%%%%%%%%%%%%%%%%%

In addition to the minimal UED action, we consider bulk mass and boundary terms for the third generation and a generic boundary term which is  parameterized as
\bea
S_{NMUED}&=&S_{MUED}+\int d^4x \int_{-L}^L dy \left\{-M_t \xi_{L/R} \overline{\Psi}_3\Psi_3 \right.\nonumber\\
&&\left. +\left[\delta(y-L)+\delta(y+L)\right] \left[r_g \mathcal{L}_{MUED}+(r_t-r_g) i \overline{\Psi}_{3,L/R}\slashed{D}\Psi_{3,L/R}\right]\right\},
\eea
where $L=\pi R/ 2$ with $R$ being the compactification radius, $\xi_{L/R}=\pm 1$, and $\Psi_3=\left\{Q_3,T,B\right\}$ are 5D fermions containing the third generation quarks, of which $Q_3$ has a left-handed and $T,B$ have right-handed boundary terms with a parameter $r_t$. 
Both bulk masses and boundary terms modify the masses and wave functions of the KK modes. A fermion  $\Psi$ with a left-handed zero mode (i.e. $Q$ and $L$) in the presence of a boundary parameter $r_t$ and a bulk mass $M_t=\mu_t\theta(y)$ is decomposed as follows.
 \begin{eqnarray}
\Psi (x,y)=\sum_{n=0}^\infty \left(\psi^{(n)}_L(x)f^{\Psi_L}_n(y)+\psi^{(n)}_R(x)f^{\Psi_R}_n(y)\right) \, ,
\label{fdecomp}
\end{eqnarray}
where the wave functions $f^{\Psi_{L/R}}_n$ are given by
\begin{eqnarray}
n=0: && f^{\Psi_L}_0=N^{\Psi}_0 e^{\mu_t |y|},\label{fzeromode}\\
\mbox{odd } n:&&\left\{\begin{array}{l}
f^{\Psi_L}_n=N^{\Psi}_n \mathfrak{sin}(k_n y) \, , \label{ffplus}\\
f^{\Psi_R}_n=N^{\Psi}_n \left(-\frac{k_n}{m_{f_n}}\mathfrak{cos}(k_n y)+\frac{\mu_t}{m_{f_n}}\theta(y) \mathfrak{sin}(k_n y)\right) \, ,
\end{array}
\right.\\
\mbox{even } n:&&\left\{\begin{array}{l}
f^{\Psi_L}_n=N^{\Psi}_n \left(\frac{k_n}{m_{f_n}} \mathfrak{cos}(k_n y)+\frac{\mu_t}{m_{f_n}} \theta(y)\mathfrak{sin}(k_n y)\right) \, ,\\
f^{\Psi_R}_n=N^{\Psi}_n \mathfrak{sin}(k_n y) \, . \label{fKKmodelast}\\
\end{array}
\right.
\end{eqnarray}
where $\mathfrak{sin}$ and $\mathfrak{cos}$ denote $\sin$ or $\sinh$ and $\cos$ or $\cosh$, and  wave numbers $k_n$ are the solutions of the mass quantization condition
\begin{eqnarray}
\begin{array}{ll}
 k_n \mathfrak{cos}(k_n L)=(r_t \left(m_{f_n}\right)^2+\mu_t)\mathfrak{sin}(k_n L)  \,  &  ~~{\rm for~ odd~ }n \, , \\
r_t k_n \mathfrak{cos}(k_n L)=-(1+r_t \mu_t)\mathfrak{sin}(k_n L) \,  & ~~  {\rm for~ even~ }n \, .
\label{fKKnumb}
\end{array}
\end{eqnarray}
The chiral zero mode is massless.
If ``light'' ($\sinh$ and $\cosh$) solutions exist, they describe the first and second KK excitations, and their masses $m_{f_n}$ is given by
\begin{eqnarray}
m_{f_n}=\sqrt{-k_n^2+\mu_t^2},
\label{fKKmassl}
\end{eqnarray}
while the ``heavy'' KK modes  ($\sin$ and $\cos$ solutions) have masses
\begin{eqnarray}
m_{f_n}=\sqrt{k_n^2+\mu_t^2}.
\label{fKKmass}
\end{eqnarray}
The normalization factors are given by 
\begin{eqnarray}
N^{\Psi}_n= \left\{
\begin{array}{lr}
\sqrt{  \frac{\mu_t}{ ( 1 +2 r_t \,  \mu_t) \exp \left ( 2 \mu_t L  \right ) - 1}} &\mbox{for } n=0\, ,\\
\frac{1}{\sqrt{L - \frac{\mathfrak{cos} (k_n L) \mathfrak{sin}(k_n L) }{k_n} + 2 r_t \mathfrak{sin}^2 ( k_n L) } }&\mbox{for odd } n\, ,\\
 \frac{1}{\sqrt{L - \frac{\mathfrak{cos} (k_n L) \mathfrak{sin}(k_n L) }{k_n}  } }&\mbox{for even } n\, ,
\end{array}
\right.
\label{Nf}
\end{eqnarray}
and they are determined from the modified orthogonality relations
 \begin{eqnarray}
\int_{-L}^L dy \, f^{\Psi_L}_m f^{\Psi_L}_n[1+r_t\left[\delta(y+L)+\delta(y-L)\right]=\delta_{mn},\nonumber\\
\int_{-L}^L dy \, f^{\Psi_R}_m f^{\Psi_R}_n=\delta_{mn}.
\label{fermionsclprd}
\end{eqnarray}
A fermion with a right-handed zero mode (i.e. $U$, $D$, $E$) yields analogous results when replacing $\mu_t$ with $-\mu_t$.  

The KK reduction of gauge bosons and scalars has been discussed in Ref. \cite{FMP}. The fields are decomposed according to
\begin{eqnarray}
A_\mu(x,y)&=&\sum_{n=0}^\infty A_\mu^{(n)}(x)f^A_n(y) \label{gaugeKKdecomp} \, ,\\
H(x,y)&=&\sum_{n=0}^\infty H^{(n)}(x)f^A_n(y) \, .
\end{eqnarray}
For a uniform boundary kinetic term as considered in this article, the resulting wave functions  are\footnote{For generic choices of the boundary parameters, the KK decomposition in the electroweak sector is more involved. For a detailed discussion and the general solutions, we refer to Ref. \cite{FMP}.}
\begin{eqnarray}
n=0: && f^A_0(y)= \frac{1}{ \sqrt{ 2 L ( 1 + \frac{r_g}{L})}}\label{gaugezm} \, , \\
\mbox{odd } n:&& f^A_n(y)= \sqrt {  \frac{1}{L+ r_g \sin^2 (k_n L)} }  \sin(k_n y)\label{gaugeom} \, , \\
\mbox{even } n:&& f^A_n(y)= \sqrt {  \frac{1}{L+ r_g \cos^2 (k_n L)} }  \cos( k_n y)\label{gaugeem} \, ,
\end{eqnarray}
where the wave numbers $k_n$ are determined by 
\begin{eqnarray}
 \label{gaugeKKmass}
 \cot (k_n L) &=  r_g k_n  \,    &~~{\rm for~ odd~ }n ,  \\\nonumber
 \tan (k_n L) &= - r_g k_n  \,   &~~  {\rm for~ even~ }n  ,
\end{eqnarray}
and the corresponding KK masses are 
\begin{equation}
m_{A_n}  = \sqrt{k^2_n+m_0^2}  \,, 
\label{gKKmass}
\end{equation}
where $m_0$ is the zero mode mass ($m_W$, $m_Z$, $m_H$ or zero), which are induced by electroweak symmetry breaking (EWSB).
The wave functions satisfy the following orthogonality relation
\begin{eqnarray}
\int_{-L}^L dy f^A_mf^A_n\left[1+r_g\left(\delta(y+L)+\delta(y-L)\right)\right]=\delta_{mn}.
\label{gaugesclprd}
\end{eqnarray}
As expected, the masses and wave functions for scalars and gauge bosons are identical to those of the $Z_2$-even fermions in the limit $\mu_t\rightarrow 0$.

The couplings are obtained from overlap integrals of the respective wave functions. As an example, the coupling of a zero mode gauge boson with KK mode fermions follows from
\begin{eqnarray}
\begin{split}
S_{eff}&\supset \int d^4x\, i g_5 \overline{\psi}^{(n)} _{L/R}\gamma^\mu A_\mu^{(0)}\psi^{(m)}_{L/R} \int^L_{-L} dy f^A_0 f^{\Psi_{L/R}}_n f^{\Psi_{L/R}}_m \left[1+r_t\left(\delta(y+L)+\delta(y-L)\right)\right]\\
&=\int d^4x\, i \frac{g_5 \delta_{nm}}{\sqrt{2L(1+r/L)}} \overline{\psi}^{(n)} _{L/R}\gamma^\mu A_\mu^{(0)}\psi^{(m)}_{L/R} \, ,
\end{split}
\end{eqnarray}
implying that
\begin{eqnarray}
g^{\rm eff}_{0nm}=\frac{g_5 \delta_{nm}}{\sqrt{2L(1+r/L)}}=g_{\rm SM}\delta_{nm} \, ,
\label{eq:geff}
\end{eqnarray}
where $g_{\rm SM}$ is the standard model coupling. All couplings of zero mode gauge bosons to KK fermions are Kaluza-Klein number conserving and of strength $g_{SM}$, {\it i.e.} independent of the fermion KK level. Calculation of the analogous overlap integrals yields the same result for couplings of zero mode gauge bosons to KK gauge bosons, and of the zero mode Higgs to KK mode fermions or gauge bosons. 
Interactions between the zero mode Higgs and KK fermions are  given by the standard model Yukawa couplings 
if we assume the same mass and boundary terms for $Q_3,U_3,$ and $D_3$, which is the case in our current study.

%%%%%%%%%%%%%%%%%%%%%%%%%%%%%%%%%%%%%%%%%%%%%%%%%%%%%%%%%%%%
\section{Higgs production and decay into photons in NMUED}
\label{sec:higgs}
%%%%%%%%%%%%%%%%%%%%%%%%%%%%%%%%%%%%%%%%%%%%%%%%%%%%%%%%%%%%

In the standard model, the production and decay processes of the Higgs boson, $gg\to H$ and $H\to \gamma\gamma$, are radiatively induced by triangle diagrams where top quark and $W$ boson are involved. In UED, there are additional contributions from the Kaluza-Klein partners amongst which KK-top and KK $W$-loops dominate. 

For a general triangle diagram, KK number is not necessarily conserved in the loop which in particular requires to sum over all allowed combinations of KK numbers in the loop when calculating the amplitude. For Higgs production and decay, however, all external lines are zero mode bosons (photon, gluon or Higgs), which preserve KK number as shown in the last section. Furthermore, as shown in Eq.~(\ref{eq:geff}), the $0$-$n$-$n$ couplings are all given by the corresponding standard model couplings. Therefore, the UED contributions to  $gg\to H$ and $H\to \gamma\gamma$ can be easily calculated from the standard model expressions ({\it c.f.} \cite{Gunion:1989we} and \cite{Belanger:2012mc}) by replacing the standard model masses with the masses of the respective KK particles given in Eqs.~(\ref{fKKmassl},\ref{fKKmass},\ref{gKKmass}) and summing over all KK modes. 
Note that the KK fermions are Dirac and therefore enter with twice the weight of the Standard model fermions.

Through triangle diagrams with W-bosons, top-quarks and their Kaluza-Klein excitations, the Higgs boson would decay to a pair of gluons and photons with the widths:
\begin{eqnarray}
\Gamma_{H\to gg}^{UED} &=& {\cal K} \frac{m_H^3}{8 \pi v^2}\cdot \frac{\alpha_S(m_H)^2}{\pi^2}|F_t|^2,\\
\Gamma_{H\to \gamma\gamma}^{UED} &=& \frac{m_H^3}{16 v^2 \pi}\cdot \frac{\alpha(m_H)^2}{\pi^2}|F_W+3Q_t^2 F_t|^2,
\end{eqnarray}
where ${\cal K} \approx 1.9$ is the QCD ${\cal K}$ factor in NNLO calculation \cite{Ravindran:2003um}, the Higgs vacuum expectation value is $v= \left(\sqrt{2} G_F\right)^{-1/2}$ with the Fermi-constant $G_F = 1.1663787(6)\times 10^{-5}\,{\rm GeV^{-2}}$ and 
$Q_t =2/3$ is electric charge of the top quark .  The one-loop functions are the sum of the SM contribution and the KK contributions: $F_W = F_W^{\rm SM} + F_W^{\rm KK},\, F_t = F_t^{\rm SM} + F_t^{\rm KK}$ with 
\begin{eqnarray}
&&F_W^{\rm SM}=\frac{1}{2} + \frac{3}{4} \tau_W +\frac{3}{2}\tau_W (1-\frac{\tau_W}{2})C_0(\tau_W), \nonumber \\
&&F_W^{\rm KK}=\sum_{n=1}^\infty \left[\frac{1}{2}+\tau_W +2 \left(\tau_W(1-\frac{\tau_{W_n}}{2})-\frac{\tau_{W_n}}{4}\right)C_0(\tau_{W_n})\right], \nonumber \\
&&F_t^{\rm SM}=-\frac{\tau_t}{2} \left[1+\left(1-\tau_t\right)C_0(\tau_t)\right], \nonumber \\
&&F_t^{\rm KK}=\sum_{n=1}^\infty \left[\frac{\tau_{t}}{\tau_{t^L_n}} F_t^{\rm SM}(\tau_t\to\tau_{t^L_n})+\frac{\tau_{t}}{\tau_{t^R_n}} F_t^{\rm SM}(\tau_t\to\tau_{t^R_n})\right],
\end{eqnarray}
where the masses are conveniently parameterized as $\tau_W ={4 m_W^2}/{m_H^2}$, $\tau_t ={4 m_t^2}/{m_H^2}$, 
$\tau_{W_n}= {4 m_{W_n}^2}/{m_H^2}$ and $\tau_{t^{L/R}_n}= (4 m_{t^{L/R}_n}^2)/({m_H^2})$.
The three point  Passarino-Veltman function is $C_0(\tau) =\left[\sin^{-1}(1/\sqrt{\tau})\right]^2$ for $\tau\geq 1$.
$F_W^{\rm KK}$ includes contributions of KK excitations in the gauge and Higgs sectors. 
The production cross section $\hat{\sigma}_{gg\to H}(\hat{s})$ is related to the partial decay width as follows, 
\begin{eqnarray}
\hat{\sigma}_{gg\to H}(\hat{s})=\frac{\pi^2}{8m_H} \Gamma_{H\to gg}(m_H)\delta(\hat{s}-m_H^2) \, .
\end{eqnarray}

%%%%%%%%%%%%%%%%%%%%%%%%%%%%%%%%%%%%%%%%%%%%%%%%%%%%%%%%%%%%
\section{Electroweak precision tests ($S,T,U$) parameters}
\label{sec:stu}
%%%%%%%%%%%%%%%%%%%%%%%%%%%%%%%%%%%%%%%%%%%%%%%%%%%%%%%%%%%%

A strong bound on the MUED model arises from UED contributions to the Peskin-Takeuchi parameters $S$, $T$, and $U$ \cite{ewbounds}, which parameterize the oblique corrections to the electroweak gauge boson propagators \cite{PT1}. The analysis of the Peskin-Takeuchi parameters in UED models with bulk masses and boundary kinetic terms has been performed in Ref. \cite{Flacke:2013pla}. The dominant contributions to $S$ and $T$ arise from the top-loop corrections to the gauge boson propagators, while $U$ only receives contributions from gauge boson and Higgs loops. At one-loop order, the UED contributions to the Peskin-Takeuchi parameters are \cite{ewbounds,Flacke:2013pla}\footnote{In Ref. \cite{Flacke:2013pla}, additional contributions to $T$ and $U$ arise due to lepton bulk mass terms, which are however shown to be strongly constraint by dilepton searches. Here, we assume vanishing lepton bulk masses and therefore neglect such contributions.}
 \begin{eqnarray}
S_{UED}&=&\frac{4 \sin^2\theta_W}{\alpha}\left[\frac{3g^2_{ew}}{4(4\pi)^2}\left(\frac{2}{9}\sum_n \frac{m^2_t}{m^2_{t^{(n)}}}\right)+\frac{g^2_{ew}}{4(4\pi)^2}\left(\frac{1}{6}\sum_n \frac{m^2_h}{m^2_{h^{(n)}}}\right)\right] \, ,  \label{Seff} \\
T_{UED}&=&\frac{1}{\alpha}\left[\frac{3g^2_{ew}}{2(4\pi)^2}\frac{m_t^2}{m_W^2}\left(\frac{2}{3}\sum_n \frac{m^2_t}{m^2_{t^{(n)}}}\right)+\frac{g^2_{ew}\sin^2\theta_W}{(4\pi)^2\cos^2\theta_W}\left(-\frac{5}{12}\sum_n\frac{m^2_h}{m^2_{h^{(n)}}}\right)\right] \, ,  \label{Teff} \\
U_{UED}&=&-\frac{4\sin^2\theta_W}{\alpha}\left[\frac{g^2_{ew}\sin^2\theta_W}{(4\pi)^2}\left(\frac{1}{6}\sum_n\frac{m^2_W}{m^2_{W^{(n)}}}-\frac{1}{15}\sum_n\frac{m^2_Wm^2_h}{m^2_{W^{(n)}}m^2_{h^{(n)}}}\right)\right] \, . \label{Ueff}
 \end{eqnarray}
Here $\alpha$ is the fine structure constant, $\theta_W$ is the Weinberg angle, and $g_{ew}$ is the coupling strength of $SU(2)_W$.

%%%%%%%%%%%%%%%%%%%%%%%%%%%%%%%%%%%%%%%%%%%%%%%%%%%%%%%%%%%%
\section{Results}
\label{sec:result}
%%%%%%%%%%%%%%%%%%%%%%%%%%%%%%%%%%%%%%%%%%%%%%%%%%%%%%%%%%%%

In our analysis, for simplicity we consider a common bulk mass $\mu_t\equiv \mu_{Q_3}=\mu_B=\mu_T$ and 
a common boundary parameter $r_t\equiv r_{Q_3}=r_B=r_T$, 
for the third generation of $SU(2)_W$ quark doublet $Q_3$ and $SU(2)_W$ singlets $B$ and $T$. 
This choice, together with the compactification scale $R^{-1}$, leaves us with three parameters. As an additional parameter, we consider a common boundary parameter $r_{g}$ for all other fields ({\it i.e.} the Higgs, the gauge fields and leptons, and the first and second family quarks) in order to illustrate how the bounds change in the presence of a common boundary parameter with only the third family quarks differing. 
We present results as bounds on the compactification scale $R^{-1}$ as a function of the dimensionless parameters $\mu_t L$ and $r_t/L$. To indicate the effect of a common boundary term, we show constraints for $r_g/  L= 0$ (``vanishing boundary parameter'') and $r_g / L=1$ (``typical boundary parameter'') \footnote{A naive dimensional analysis of the boundary parameter yields $r/L\lesssim 12/ \Lambda R$, where $\Lambda$ is the UED cutoff scale, and $\Lambda R$ gives an estimate for the number of KK levels below the cutoff scale \cite{CTW}.}. 

\begin{figure}[t]
\centering
\includegraphics[width=0.47\textwidth]{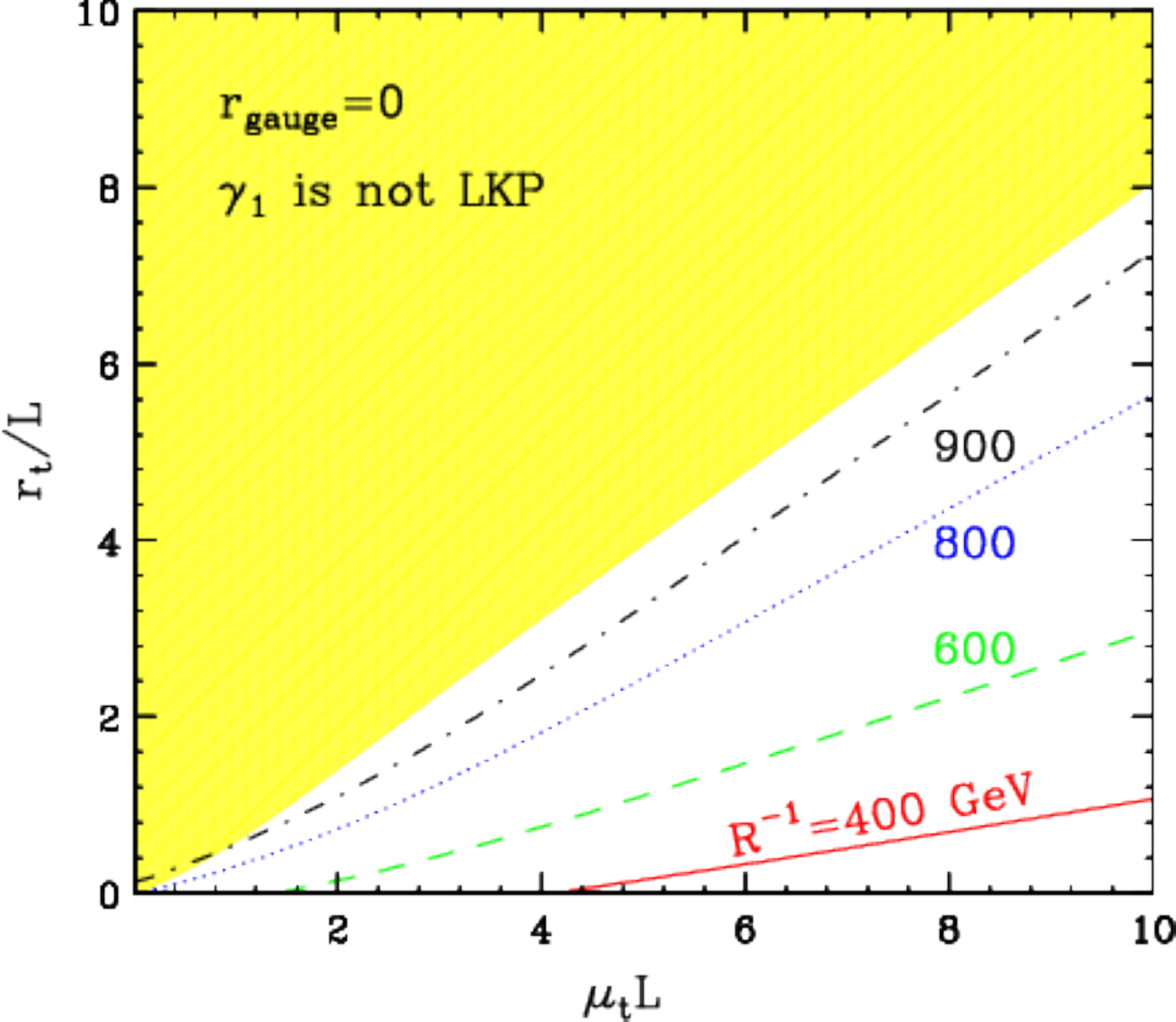} \hspace{0.3cm}
\includegraphics[width=0.47\textwidth]{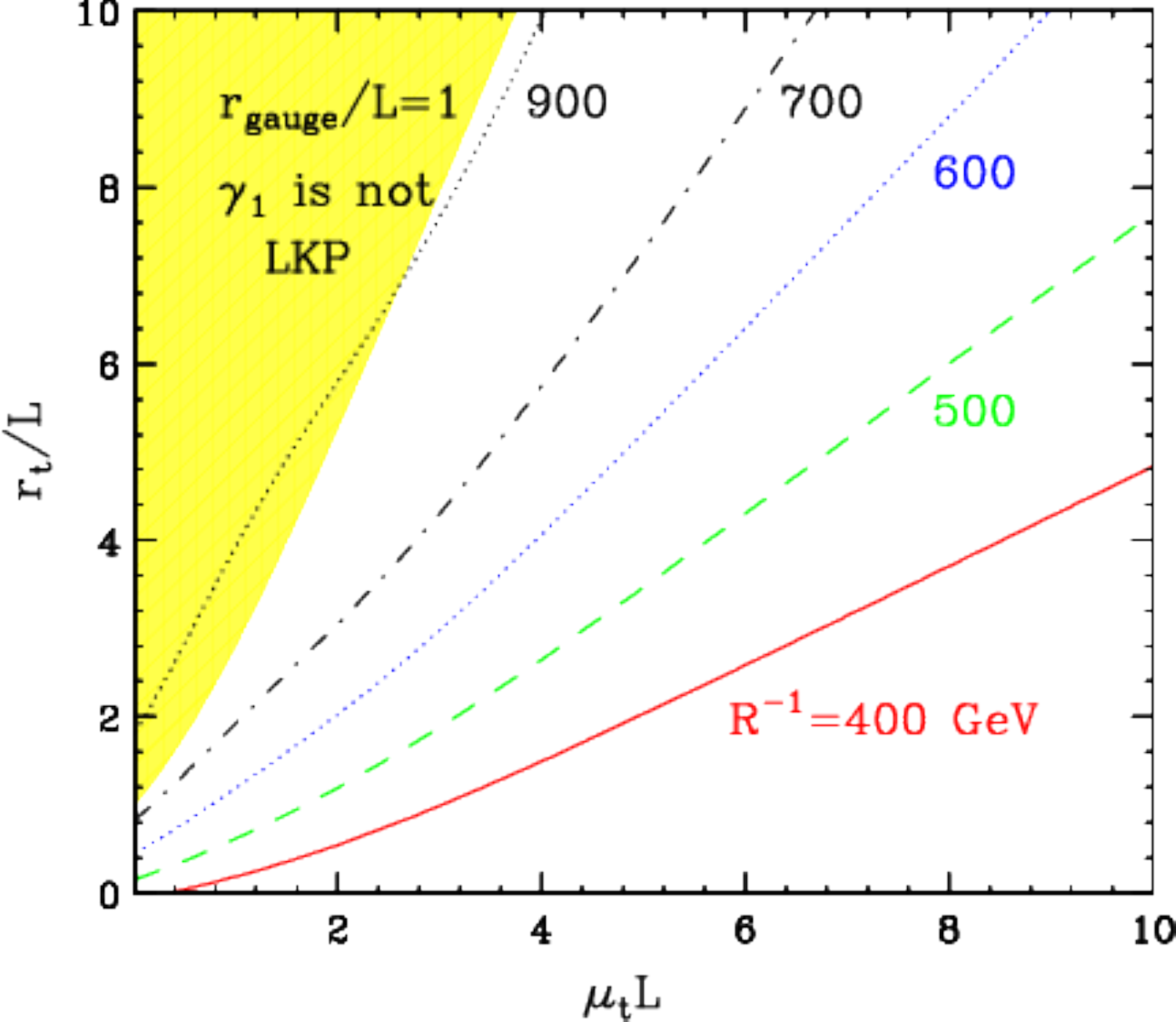}
\caption{Oblique ($S,T,U$) bounds on $R^{-1}$ with the latest Gfitter data with $m_H=126$ GeV.  The plots show contours of minimally allowed $R^{-1}$ in the $r_t/L$ vs. $\mu_t L$ parameter space for $r_g/L=0$ (left) and $r_g/L=1$ (right). 
The shaded region (in yellow) is excluded since KK bottom is the lightest Kaluza-Klein particle.}
\label{fig:EW}
\end{figure}

The electroweak bounds shown in Fig.~\ref{fig:EW} are obtained by performing a $\chi^2$ fit of the parameters $S_{UED}$, 
$T_{UED}$, $U_{UED}$ from Eqs. (\ref{Seff})-(\ref{Ueff}) to the experimental values  given in Ref.~\cite{ewGfitter}, 
$S_{NP}=0.03\pm0.10\, ,\, T_{NP}=0.05\pm0.12\, , \, U_{NP}=0.03\pm0.10$, 
for a reference point $m_{H}=126 ~\mbox{GeV}$  and $m_{t}=173 ~ \mbox{GeV}$ with  correlation coefficients  of $+0.89$ between $S_{NP}$ and $T_{NP}$, and $-0.54$  $(-0.83)$ between $S_{NP}$ and $U_{NP}$ ($T_{NP}$ and $U_{NP}$). 

For  $r_g/L=0$, the mass of the first $U(1)_Y$ KK mode $\gamma^{(1)}$ (the usual UED dark matter candidate) is given by $R^{-1}$. 
For a large $r_t/L$  and a small $\mu_t/L$, the first KK bottom partner is lighter than the $\gamma^{(1)}$ which implies a charged dark matter and is therefore excluded. 
For  $r_g/L\neq 0$, the same applies, although the mass of the $\gamma^{(1)}$ is not  given by $R^{-1}$ anymore, but determined by Eq.~(\ref{gaugeKKmass}).

\begin{figure}[t]
\centering
\includegraphics[width=0.47\textwidth]{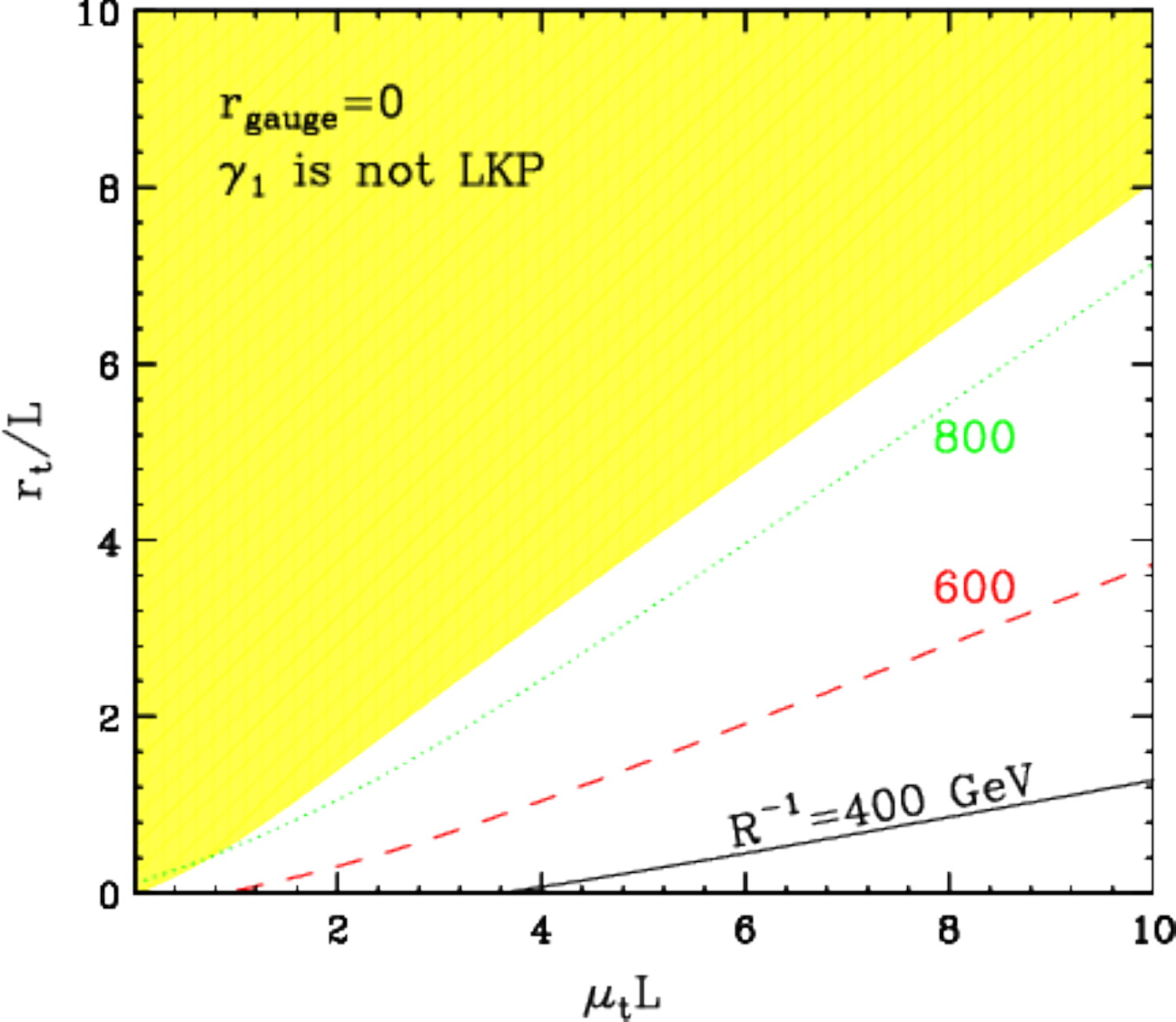} \hspace{0.3cm}
\includegraphics[width=0.47\textwidth]{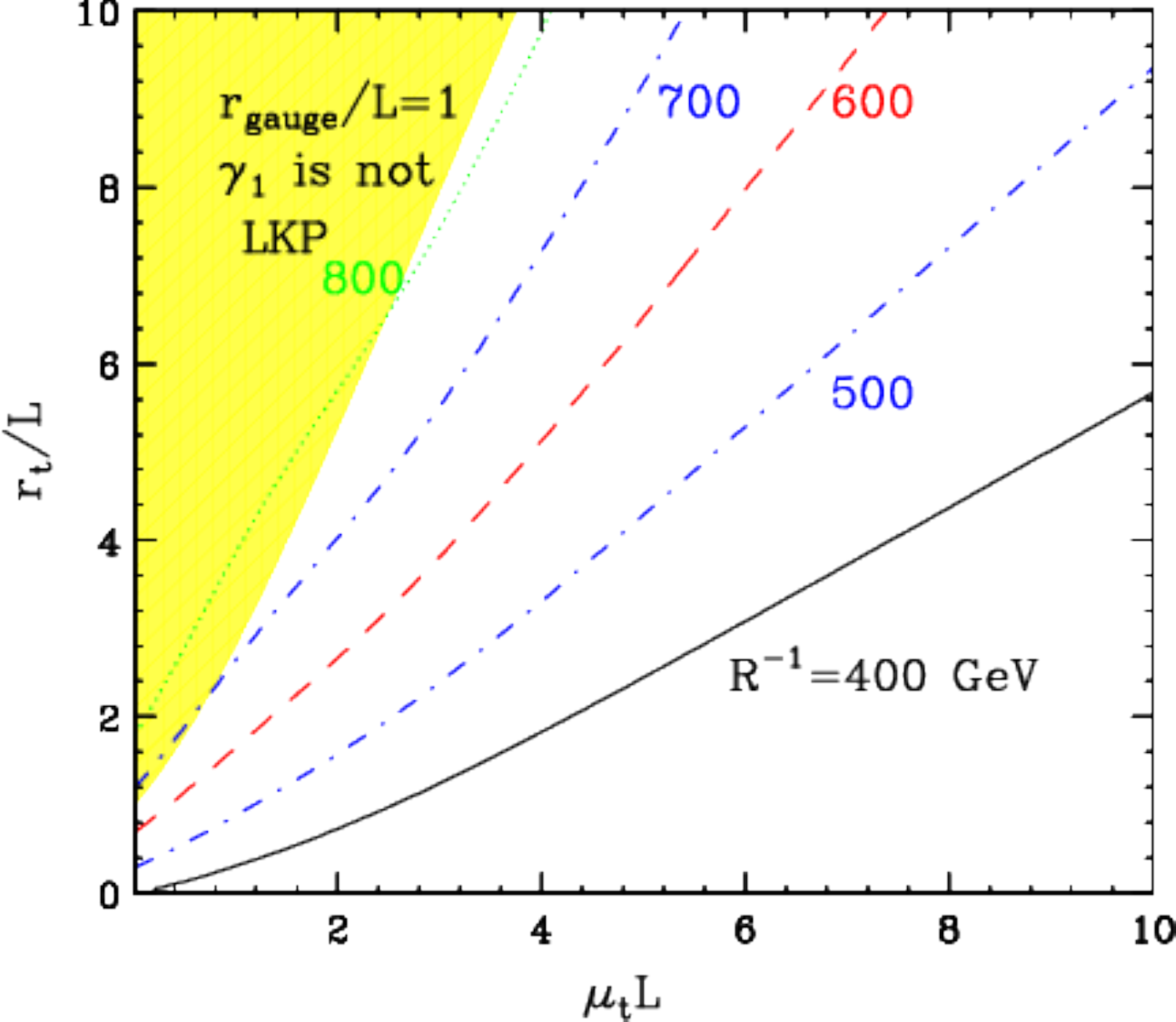} 
\caption{Combined constraints from Higgs searches at ATLAS and CMS. The plots show contours of minimally allowed $R^{-1}$ in the $r_t/L$ vs. $\mu_t/L$ parameter space for $r_g/L=0$ (left) and $r_g/L=1$ (right). 
The shaded region (in yellow) is excluded since Kaluza-Klein photon is not a dark matter  candidate.}
\label{fig:Comb}
\end{figure}

To determine the bounds from Higgs searches, we define the signal strengths as follows: 
\bea
\mu_{gg\rightarrow h \rightarrow \gamma \gamma}&\equiv& \frac{\hat{\sigma}^{UED}_{gg\rightarrow h \rightarrow \gamma \gamma}}{\hat{\sigma}^{SM}_{gg\rightarrow h \rightarrow \gamma \gamma}}=\frac{|F_t|^2|F_W+3Q_t^2 F_t|^2}{|F^{SM}_t|^2|F^{SM}_W+3Q_t^2 F^{SM}_t|^2}, \\
\mu_{other \rightarrow h \rightarrow \gamma \gamma}&\equiv& \frac{\hat{\sigma}^{UED}_{other\rightarrow h \rightarrow \gamma \gamma}}{\hat{\sigma}^{SM}_{other\rightarrow h \rightarrow \gamma \gamma}}=\frac{|F_W+3Q_t^2 F_t|^2}{|F^{SM}_W+3Q_t^2 F^{SM}_t|^2},  \\
\mu_{gg\rightarrow h \rightarrow other}&\equiv& \frac{\hat{\sigma}^{UED}_{gg\rightarrow h \rightarrow other}}{\hat{\sigma}^{SM}_{gg\rightarrow h \rightarrow other}}=\frac{|F_t|^2}{|F^{SM}_t|^2},  \\
\mu_{other \rightarrow h \rightarrow other}&\equiv& \frac{\hat{\sigma}^{UED}_{other\rightarrow h \rightarrow other}}{\hat{\sigma}^{SM}_{other\rightarrow h \rightarrow other}}=1, \label{mudefs} 
\eea
where ``other'' production channels are vector boson fusion, and Higgs radiation off gauge bosons or tops, and ``other'' decay channels are $ZZ$, $WW$, $b\bar{b}$, and $\tau\tau$. Ref. \cite{Dumont:2013wma} performs a global Bayesian analysis on the ATLAS \cite{ATLASrefs} and CMS \cite{CMSrefs} Higgs data and provides values of signal strengths and their correlations. 
We use this data to perform a $\chi^2$ test of the UED predictions and plot the constraints in Fig. \ref{fig:Comb} where no correlation between the ATLAS and CMS results is assumed.

Compared to the electroweak constraints and the CMS bounds, the ATLAS bounds are weaker. The main reason for this lies in that ATLAS (CMS) observes an enhanced (reduced) rate in the di-photon channel as compared to the standard model expectation. 
The UED model predicts an enhancement of the cross section for $gg\rightarrow h \rightarrow \gamma \gamma$.\footnote{To be more precise, the decay rate $h\rightarrow \gamma\gamma$ is reduced due to negative interference of the KK top and KK $W$ loop contributions with the Standard Model contribution, but the production cross-section $gg\rightarrow h$ is enhanced and leads to an enhancement of the overall cross section.  The UED cross sections for $gg\rightarrow h\rightarrow other$ is enhanced even more, which again is partially reflected in some search channels at ATLAS, but not at CMS.}
The other channels do not have a strong effect on the $\chi^2$ fit because of the larger errors. Therefore, the constraints shown in Fig. \ref{fig:Comb} are dominated by the measurements at CMS. We note that the bounds from electroweak precision tests (with $m_h=126$~GeV) and from direct Higgs measurements are comparable or slightly weaker.

%

%%%%%%%%%%%%%%%%%%%%%%%%%%%%%%%%%%%%%%%%%%%%%%%%%%%%%%%%%%%%
\section{Summary and outlook}
\label{sec:conclusion}
%%%%%%%%%%%%%%%%%%%%%%%%%%%%%%%%%%%%%%%%%%%%%%%%%%%%%%%%%%%%

UED is an attractive extension of the standard model based on higher dimensions providing a viable dark matter candidate and rich phenomenology at the LHC. As an effective theory, UED models could be extended from the minimal realization by incorporating boundary localized operators and bulk masses. In this paper, we focus on the extra terms associated with the third generation of quarks, which are particularly sensitive to the radiative production and decay of the Higgs boson through the Kaluza-Klein quarks and $W$-bosons. 
Including the electroweak precision tests as well as the latest measurements on the Higgs boson at the LHC (ATLAS and CMS), 
we explicitly show the allowed range of parameter space in Next-to-Minimal UED for the Kaluza-Klein photon dark matter candidate.
Our results show that NMUED allows for a lower compactification scale than as in MUED, where 
$R^{-1} < 500$ GeV is excluded at 95\% C.L. \cite{Belanger:2012mc}.\footnote{The analysis in in Ref. \cite{Belanger:2012mc} is based on earlier ATLAS and CMS Higgs data. With the data set used here, these bounds are expected to increase. In the $0=\mu_t=r_t=r_g$ limit of our analysis, we find $R^{-1} \lesssim 700$~GeV excluded. To our knowledge, the currently strongest published MUED bound from other collider searches is $R^{-1} < 715$~GeV \cite{Edelhauser:2013lia}.} 
This allowed parameter space will be probed by the LHC14.

%\bigskip
%%%%%%%%%%%%%%%%%%%%%%%%%%%%%%%%%%%%%%
%\section*{Acknowledgments}
\acknowledgments
%%%%%%%%%%%%%%%%%%%%%%%%%%%%%%%%%%%%%%
TF is supported by the National Research Foundation of Korea (NRF) grant funded by the Korea government (MEST) N01120547. 
KK is supported in part by the US DOE Grant DE-FG02-12ER41809 and by the University of Kansas General Research Fund allocation 2301566. 
SC is supported by Basic Science Research Program through the National Research Foundation of Korea 
funded by the Ministry of Education, Science and Technology (2011-0010294) and (2011-0029758).

%%%%%%%%%%%%%%%%%%%%%%%%%%%%%%%%%%%%%%%%%%%%%%%%%%%%%%%%%%%%

\end{document}